\documentclass[12pt]{article}

\usepackage{amsmath,amssymb}
\usepackage{amsthm}

\newcommand{\be}{\begin{equation}}
\newcommand{\ee}{\end{equation}}
\newcommand{\bea}{\begin{eqnarray}}
\newcommand{\eea}{\end{eqnarray}}

\newcommand{\gapp}{\mathrel{\raise.3ex\hbox{$>$}\mkern-14mugo
              \lower0.6ex\hbox{$\sim$}}}
\newcommand{\lapp}{\mathrel{\raise.3ex\hbox{$<$}\mkern-14mu
              \lower0.6ex\hbox{$\sim$}}}

\newcommand\lsim{\lesssim}

\newcommand\gsim{\gtrsim}

\renewcommand\({\left(}
\renewcommand\){\right)}

\newcommand\eq[1]{Eq.~(\ref{#1})}

\newcommand\mpl{M_{\rm P}}

\newcommand\dbibitem[1]{\bibitem{#1}}
\newcommand{\dlabel}[1]{\label{#1}}

\def\calp{{\cal P}}

\def\calpz{\calp_\zeta}

\newcommand\GeV{\,\mbox{GeV}}
\newcommand\MeV{\,\mbox{MeV}}

\newcommand\Mpc{\,\mbox{Mpc}}

\newcommand\sub[1]{_{\rm #1}}

\newcommand\mone{^{-1}}
\newcommand\mtwo{^{-2}}

\newcommand\half{^{1/2}}

\newcommand\quarter{^{1/4}}

\newcommand\mn{{\mu\nu}}

\newcommand{\fnl}{f\sub{NL}}

\newcommand{\xls}{x\sub{ls}}

\newcommand{\zetai}{\zeta\sub{inf}}
\newcommand{\calpi}{\calp_{\zetai} }
\newcommand{\ns}{n\sub s}
\newcommand{\alphas}{\alpha\sub s}

\begin{document}

\title{BICEP2, the curvature perturbation and supersymmetry}
\author{David H.\ Lyth\\Consortium for Fundamental Physics,\\ Cosmology and
Astroparticle Group, Department of Physics,\\ Lancaster University,
Lancaster LA1 4YB, UK}
\maketitle
\begin{abstract}
The  tensor fraction $r\simeq 0.16$ found by  BICEP2  corresponds to a Hubble parameter
$H\simeq 1.0\times 10^{14}\GeV$ during inflation.
 This has two implications for the (single-field) slow-roll inflation
hypothesis. First,  the inflaton perturbation
must account for much more than $10\%$ of the  curvature perturbation $\zeta$, which barring fine-tuning means that it
accounts for practically all of it. It follows that a curvaton-like mechanism  for generating $\zeta$ requires
an alternative to slow roll such as k-inflation. Second, accepting slow-roll inflation,    the excursion of  the inflaton field
 is at least of order Planck scale. As a result, the flatness of the inflaton presumably requires a shift symmetry.
 I point out that if such is the case, the resulting potential is likely to have  at least approximately  the quadratic form suggested
 in 1983 by Linde, which is known to be compatible with the observed $r$ as well as the observed spectral index $\ns$.
 The shift symmetry does not require supersymmetry. Also, the big $H$ may rule out a
   GUT by restoring the symmetry and producing fatal cosmic strings. The absence of a GUT would correspond to the
   absence of superpartners for the Standard Model particles, which indeed have yet to be found at the LHC.
 \end{abstract}

Recently, BICEP2 \cite{bicep2} has detected primordial gravitational waves. In this note I discuss two consequences of their result for
the generation of the curvature perturbation, and some of their implications.

The BICEP2 measurements gives  (after subtracting an estimated foreground) $r=0.16^{+0.06}_{-0.05}$ where $r= \calp\sub {ten}(k)/\calpz(k)$ is the tensor fraction,
evaluated on the scales  $\xls/100 \lsim k\mone \lsim \xls$ and
 $\xls=14,000\Mpc$ is the distance to the last-scattering surface.
(I will call  these large scales).
The quoted uncertainty  is only statistical and it may be that the foreground
(dust) accounts for all of the observed polarization. In this paper I discount that and accept that a tensor perturbation has been detected. Also, I will assume that the  tensor perturbation  is generated during inflation
with Einstein gravity; then  $\calp\sub{ten}(k)=(8/\mpl^2)(H(k)/2\pi)^2$
where $H(k)\equiv \dot a/a$ is the Hubble parameter at the epoch of horizon exit $k=aH$.

The spectrum $\calpz(k)$
of the curvature perturbation is measured accurately \cite{planck} on scales
$10\Mpc \lsim k\mone \lsim \xls$. (I will call these cosmological scales.)
It is nearly constant and equal to
   $\calpz\half(k_0)=4.69\pm 0.02$ at a `pivot scale
  $k_0\mone =20 \Mpc$.
     With $r=0.16$ this corresponds to
$H=1.0\times 10^{14}\GeV$, where $H$ without an argument is the large-scale value.
This corresponds to an energy scale $\rho\quarter=1.5\times 10^{16}\GeV$ which is not far below the Planck scale
$\mpl=2.4\times 10^{18}\GeV$. With my assumption, four-dimensional field theory is valid up to this scale and the string
theory from which it may be derived is relevant only at higher scales even closer to the Planck scale.

 For $\ns(k)-1\equiv d\ln \calpz/d\ln k$ observation requires
$\ns(k_0)-1=0.039\pm 0.005$ assuming $\alphas\equiv
\ns'(k_0)=0$.
 Assuming just constant $\alphas$
doesn't change the result for $\ns-1$ much, and gives
$\alphas(k_0)=-0.014^{+0.016}_{-0.017}$ at $95\%$ confidence.\footnote
{Except where stated all the  other uncertainties are  at  $68\%$ confidence levels.}
These results  suggest that
 $\ns(k)-1$ is nearly constant on cosmological scales.\footnote
 {A significant change in $\ns(k)$ has been suggested \cite{bicep2,gong}, as one way of resolving  the tension between the BICEP2 result
 and the Planck result \cite{planck} $r<0.11$ ($95\%$ confidence).
 When I come to consider models of slow-roll inflation I will assume that the tension is resolved in some other way.}

 On the usual assumption that $H$ is nearly
constant throughout inflation, the measured value of $H$  determines the  number of $e$-folds of inflation after
the scale $k = 1/\xls$ leaves the horizon given an assumed evolution of the scale factor after inflation.
Assuming matter domination until reheating at temperature $T\sub R$, it is
\be
N(1/\xls) = 61- \frac13 \ln \frac{10^9\GeV}{T\sub R}
 .\ee
 Requiring only  successful BBN one could have  $T\sub R\sim 1\MeV$ which would give $N=47$, but a value
 closer to 61 is far more likely.

\subsection*{Slow-roll inflation}

 Slow-roll inflation requires
\be
\epsilon(k)\ll 1,\qquad |\eta(k)| \ll 1,
,\ee
where
  $\epsilon(k)=\mpl^2(V'/V)^2/2$ and $\eta(k)=V''/V$ evaluated at horizon exit, with $V(\phi)$ is the inflaton potential.\footnote
  {I consider only single-field inflation, excluding for example the two-field scenario of \cite{testing}.}
It  also gives
$V(N)=3\mpl^2H^2(N)$ to good accuracy, and
\be
H\mone(N) dH/dN = \epsilon,\qquad  \epsilon\mone d\epsilon/dN=2\eta-4\epsilon
,\ee
where $N(k)$ is the number of $e$-folds of inflation after $k$ leaves the horizon.
 Therefore,
 $H(N)$ and $\epsilon(N)$ hardly change while $\Delta N=1$.

Also, slow-roll gives $\ns(k)-1=2\eta(k)- 6\epsilon(k)$, and $r=16\epsilon$ where
$\epsilon$ without an  argument is the large-scale value.
Finally,
\be
d\phi/dN= \mpl^2V'/V =\pm
\sqrt{2\epsilon(N)}\mpl
. \dlabel{dphidn}\ee
Integrating this expression gives
\be
N(\phi) =\frac1{\mpl^2}\int^\phi_{\phi\sub{end}} \frac V{V'} d\phi
, \ee
where $\phi\sub{end}$ is $\phi$ at the end of inflation. For a non-hybrid model inflation ends when slow-roll fails,
ie.\ when $\max\{\epsilon,|\eta|\}\sim 1$.

\subsection*{The inflaton contribution to $\zeta$}

The first implication of $r=0.16$
for slow-roll inflation
 concerns the contribution $\zetai$ of the inflaton perturbation to
primordial curvature
perturbation $\zeta$ \cite{liberated}.\footnote
{Generic $r$ is considered \cite{liberated}, which states a converse equivalent of the present result; that
$\calpi\ll \calpz$ is equivalent to $r\ll 0.1$.}
 To be more precise, it concerns $\calpi\half(k)$ which is the rms contribution of  $\zetai$ per unit interval of $\ln k$.
 On large scales slow-roll inflation gives
\be
\frac{\calpi\half(k)}{\calpz\half(k)} = \sqrt\frac{r}{16\epsilon} \gg \sqrt\frac r{16} = 0.10
.\ee
It follows that {\em $\zetai$ accounts for much more than $10\%$ of $\zeta$, at least on large scales.}

The requirement $\zetai=\zeta$ was
taken to be essential for almost twenty years after the advent of inflation.
Then it was realised that all \cite{ourcurv} or some \cite{moroi} of  $\zeta$ could instead be generated by a curvaton field.\footnote
{See also \cite{karicurv}
for a related scenario involving a bouncing universe.}
The curvaton acquires its perturbation during
inflation but generates $\calpz$ only at some epoch after inflation when it comes to account for a significant fraction
of the energy density. Soon after it was realised  that $\zeta$ could also be generated by what one might
call a modulon field, whose perturbation modulates some process after inflation that would take place anyway  even if
the modulon didn't exist. The original proposal \cite{modulon} was to modulate the decay of the inflaton, but many other possibilities
have since been considered.

The high value of $r$ found by BICEP2 means that a  curvaton or modulon mechanism is practically incompatible
with slow-roll inflation, and completely incompatible if the mechanism accounts for practically all of $\zeta$.\footnote
{This is verified in \cite{andrew} for the linear curvaton model.}
 That is not a problem for these mechanisms though, because they assume nothing about
the mechanism of inflation. Their only input from inflation is the Hubble parameter $H(N)$ while scales of interest are leaving the horizon.
Alternatives to slow-roll are known to exist, such as k-inflation \cite{kinflation} with sound speed bigger than $c$,
which generate a contribution to $\zeta$ that is much smaller
than $\zetai$ allowing a curvaton or modulon scenario.

The inflaton scenario is distinguished from the curvaton and modulon scenarios by the non-gaussianity that it generates.
  The reduced bi-spectrum $\fnl(k_1,k_2,k_3)$ for the inflaton scenario has is  a very specific shape \cite{maldacena}
   which would be a smoking gun
  for slow-roll inflation,
but is only of order $10\mtwo$ which is probably too small ever to detect. The curvaton and modulon scenarios
in contrast generally make $\fnl$ constant, and typically of order 1 or bigger which should eventually be detectable.
Also, when $\fnl$ is constant there is a further prediction $\tau\sub{NL} = (6\fnl/5)^2$ for the trispectrum parameter $\tau\sub{NL}$.
If the curvaton or modulon evolves non-linearly $\fnl$ need not be constant \cite{bntw}
but is still typically $\gsim 1$.

Now we know that $r\simeq 0.16$,  {\em a future detection of $|\fnl(k_1,k_2,k_3)|\gg 10\mtwo$
would strongly suggest that  inflation is not slow-roll}.
Before, it would just have ruled out  $\zeta\simeq \zetai$.
Furthermore {\em an accurate verification of $\tau\sub{NL}=(6\fnl/5)^2$
would completely rule out slow-roll inflation.} Before, it would just have required that $\zetai$ is negligible.

Existing studies of curvaton and modulon scenarios leave $H(N)$ arbitrary. Now that we know $H$ on large scales the predictions are much sharper
and should all be revisited. For instance, the axionic curvaton model studied in \cite{axionic} gives, with the observed $H$
and adopting the simplest version, $\fnl \sim 1$.\footnote
{One sees this from Figure 1 of the paper.} The axionic curvaton model is particularly attractive because (i) the flatness of the potential
can protected by a shift symmetry as discussed below for the inflaton potential and (ii) it can generated the observed spectral index
even if $\epsilon_H\equiv |dH/dN|/H$ is very small.\footnote
{A curvaton or modulon model  gives  $\ns=1-2\epsilon_H+ (V''/3H^2)$ where $V$ is the curvaton potential.}

\subsection*{The big change in the inflaton field}

The second implication of $r\simeq 0.16$  for slow-roll inflation \cite{mytensor,bl}  is a lower bound on the change in the inflaton field $\phi$
during inflation.\footnote
 {In \cite{mytensor,bl}, generic $r$ was considered.
 In \cite{mytensor} an
old definition of $r$ was used,  which is $6.9/8$ times the now-standard definition.}
It comes from the relation
\eq{dphidn}.
 While large scales leave the horizon the change
in $N$ is only $\Delta N\simeq 4$. During this era $\epsilon \geq r/16$, and
 the corresponding change in $\phi$ is $\Delta_4\phi\simeq =4\sqrt{2\epsilon}$. It follows that
 $\Delta\phi$, the total change in $\phi$ after the scale $\xls$ leaves the horizon, satisfies
 \cite{mytensor}
\be
\Delta\phi\gsim 4\sqrt{r/8}\mpl = 0.56\mpl
. \dlabel{lyth1} \ee
In \cite{bl} it was pointed out that a stronger result holds  for the total change in $\phi$ if $\epsilon(k)$
doesn't decrease during inflation:
\be
\Delta\phi \geq N \sqrt{r/8} =8.4[N(1/\xls)/60] \mpl
\dlabel{lyth2}, \ee
where $N\equiv N(1/\xls)$ is the total number of $e$-folds.\footnote
{The result \eq{lyth1} (for generic $r$) is usually called the Lyth bound. Sometimes (for instance in \cite{anupam,mayang,evading})
the result
\eq{lyth2}  is also called the Lyth bound even though it was obtained by two authors  \cite{bl}).
This is done in \cite{anupam,evading}, where  potentials with decreasing $\epsilon$
 are incorrectly  said to violate
the finding of  \cite{mytensor}.}

\subsection*{The potential of slow-roll inflation}

I take the inflaton field $\phi$ to be canonically normalized, and to be described by an effective field theory valid up to some
scale $M\lsim \mpl$ (but bigger than the inflationary energy scale). One might identify $M$ with the string scale.

According to a  commonly held view, the  tree-level potential will contain all terms allowed by the symmetries.
Assuming just
  symmetry under $\phi\to -\phi$) we will have
\be
V=\frac12m^2\phi^2 +  \sum \lambda_d \phi^{d}/M^{d-4}
,\dlabel{powers} \ee
with the sum over even $d>2$. Also it is commonly assumed that in the absence of a suitably broken  symmetry
under $\phi\to \phi+$const (shift symmetry)
one will have $|m^2|\sim
M^2$ and $|\lambda_d|\sim 1$. But \eq{powers} gives
\be
\eta(N) = \frac{m^2}{3H^2} + \frac{\sum d(d-1)\lambda_d \phi^d/M^{d-4} }{ 3H^2\phi^2 }
, \ee
which barring cancellations  requires $m^2\ll 3H^2$ and
\be
d(d-1) |\lambda_d|\ll \frac{3H^2}{M^2} \( \frac M \phi \)^{d-2} \lsim
\frac{ 3H^2}{\mpl^2} \( \frac M \mpl \)^{d-4} \lsim 6\times 10^{-9}.
 \ee

One may of course dissent from the common view \cite{toni1}, but many  authors have
 taken it on board and have proposed a shift symmetry to ensure the flatness of the potential.
 Confining ourselves to the large-field case that is relevant here, three
 of the proposals  \cite{ignoble,dante,mflation} produce an approximately quadratic potential, $V\propto\phi^2$.\footnote
{ $N$-flation \cite{nflation} can do the same thing, but not for generic initial conditions even assuming equal masses
 \cite{nflation2}. I thank C.\ Gordon for pointing this out.}
Two more  \cite{extranatural,completing} produce a sinusoidal potential, corresponding to what has been
called Natural Inflation \cite{natural}. With $\phi=0$ taken to be a minimum, that too can give an approximately quadratic potential.
There is also the `monodromy' scheme \cite{monodromy} that typically gives a potential $\propto \phi^p$ with $p<2$.

A quadratic potential was suggested by Linde in 1983  \cite{chaotic},
and it accounts for the measured values of both $r$ and $\ns(k)$. To be precise it gives
\be
r=0.16 (50/N(1/\xls),\qquad n(k)-1= -0.04(50/N(k))
.\ee
with  $N(k_0)=N(1/\xls)-6.5$.
These slow-roll predictions are compatible with current observations. They can be altered slightly while maintaining agreement, by using the
 sinusoidal potential \cite{natural2}, by allowing all terms up to quadratic \cite{polynomial} or by using the $\phi^p$ potential of \cite{monodromy}.

 One can also move away from strict slow roll in at least two ways. First, in the schemes of \cite{ignoble,monodromy}
 the potential can have an additional oscillating component. That would give $\calpz(k)$ an oscillating component,
 and also give a possibly observable $\fnl$ with a distinctive shape \cite{cel}.
 Second,  the sinusoidal potential makes a
 coupling
 $\propto \phi F_\mn \tilde F^\mn$ to a gauge field  quite likely, which could have a variety of effects \cite{areview},

One can also  consider  potentials that are completely different from $\phi^2$ potential, yet give
values for $r$ and $n$ that are agreement with observation \cite{anupam2}.\footnote
{For at least the second of these, inflation after the observable universe leaves the horizon takes place with a nearly
$\phi^2$ potential. I thank Qaisar Shafi for pointing this out to me.}
  It is not clear though, how such potentials can
be the result of a broken shift symmetry.

\subsection*{Supersymmetry}

The inflationary energy density scale, $\rho\quarter = 1.5\times 10^{16}\GeV$ is the same as the GUT scale $M\sub G$,
 That must be a coincidence though, because $M\sub G$ represents the vev of the GUT Higgs fields
and not the height of their potential. The height of their potential will be some coupling $\lambda\ll 1$
times $M\sub G^4$. The energy scale of GUT inflation models, which generate the inflationary energy density
from the GUT Higgs fields, is therefore too low \cite{qaisar} to generate the observed $r$.

The high inflationary energy density is actually dangerous for  a  GUT, because it  breaks supersymmetry.
This will generate contributions typically of order $\pm H^2\simeq \pm (10^{14}\GeV)^2$ to the masses-squared of the GUT Higgs fields,
which may be bigger than their true masses-squared of order $-\lambda M\sub{GUT}^2$. In that case, if the generated contributions are positive
for at least some of the GUT Higgs field, the GUT symmetry may be at least partially restored during inflation which could produce
cosmic strings at the end of inflation that could be forbidden by observation.
The idea of a supersymmetric GUT is also endangered  by the failure so far of the LHC to find supersymmetric partners for the
Standard Model particles.

Of course, these  considerations do not rule out a GUT. One can suppose instead that the Standard Model is embedded in
split supersymmetry, and that
inflation generates GUT Higgs
mass-squared no bigger than $H$ leading to  possibly observable non-gaussianity \cite{testing}.

Even if the Standard Model has no supersymmetric partners so that there  is no GUT,
  there could still be supersymmetry broken at a high scale.
One might therefore consider supersymmetry as a mechanism for keeping the inflationary potential sufficiently flat.
  For inflation models with $\phi\ll \mpl$ supersymmetry
is indeed an attractive way of obtaining the shift symmetry, because in such models the terms with $d>2$ can be suppressed by
the factor $(\phi/\mpl)^{d-2}$ leaving only the terms $m^2\phi^2/$ and  $\lambda_4\phi^4$ to worry about.
The suppression of $\lambda_4$ can be achieved by taking the inflaton to be a flat direction of supersymmetry.
 Supergravity, in which supersymmetry will generally be embedded,  generically makes $|m^2|\sim H^2$ (the $\eta$ problem \cite{etaproblem}) but that can be solved by accepting an
order 1 percent fine tuning, or \cite{ewan} by the imposition of an additional symmetry.

Supersymmetry is far less attractive now that we need $\phi\gsim\mpl$, because
all of the $\lambda_d$ need to be suppressed.
One can achieve this by imposing an exact shift symmetry on the Kahler potential
which could give the $\phi^2$ potential \cite{chaoticsugra,fkr,renata}
but one may question the validity of doing that  \cite{fkr,chaoticsugra2} and once it is abandoned
the situation for the Kahler potential looks no better than that for the potential itself.
One could also obtain the $\phi^2$ potential from $D$-term inflation, by imposing an exact shift
symmetry on the gauge kinetic function \cite{dterm} but that too seems to have no justification.\\ \\

NOTE ADDED, May 9th 2014
The preceding text is identical with the one posted on March 13th except for the addition of a footnote 12.
Many papers have appeared since, that relate to the BICEP2 measurement. One of them \cite{jihn2} describes a
supersymmetric inflation model, which has a mechanism for suppressing the coefficients $\lambda_d$
and gives a potential which is not $\phi^2$ yet still fits observation. Another paper
\cite{last} takes the view that the equality of $\rho\quarter$ with the (supersymmetric) GUT scale is evidence for
a GUT. The idea presumably would be to identify $\rho\quarter$ with the height of the GUT Higgs potential. That would need
 $\lambda\quarter=1$ which  contradicts  the requirement $\lambda\ll 1$. It has been pointed out to me by F.~Wilczek that this requirement might
 not be essential, since the perturbative regime may be more like  $\lambda\ll 4\pi$.
The fact remains though \cite{qaisar}, that all known inflation models which {\em do} identify $\rho\quarter$ with the height of a supersymmetric
GUT Higgs potential seem to give $r\ll 0.1$, in the regime of parameter space that reproduces the observed $\calpz(k)$.

\end{document}